\documentclass[a4paper]{article}

\usepackage{INTERSPEECH2022}
\usepackage{makecell}
\usepackage{multirow}
\usepackage[hidelinks]{hyperref}

\title{Personalized Acoustic Echo Cancellation for Full-duplex Communications}
\name{Shimin Zhang$^{1}$, Ziteng Wang, Yukai Ju$^{1}$, Yihui Fu$^{1}$, Yueyue Na, Qiang Fu, Lei Xie$^{1*}$
\thanks{$^*$: Corresponding author.}
}

\address{
	$^{1}$Audio, Speech and Language Processing Group (ASLP@NPU), School of Computer Science\\
	Northwestern Polytechnical University, Xi'an, China
}
\email{shmzhang@npu-aslp.org, lxie@nwpu.edu.cn}

\begin{document}

\maketitle

\begin{abstract}
Deep neural networks (DNNs) have shown promising results for acoustic echo cancellation (AEC). But the DNN-based AEC models let through all near-end speakers including the interfering speech. In light of recent studies on personalized speech enhancement, we investigate the feasibility of personalized acoustic echo cancellation (PAEC) in this paper for full-duplex communications, where background noise and interfering speakers may coexist with acoustic echoes. Specifically, we first propose a novel backbone neural network termed as gated temporal convolutional neural network (GTCNN) that outperforms state-of-the-art AEC models in performance. Speaker embeddings like d-vectors are further adopted as auxiliary information to guide the GTCNN to focus on the target speaker. A special case in PAEC is that speech snippets of both parties on the call are enrolled. Experimental results show that auxiliary information from either the near-end speaker or the far-end speaker can improve the DNN-based AEC performance. Nevertheless, there is still much room for improvement in the utilization of the finite-dimensional speaker embeddings.

\end{abstract}

\section{Introduction}

In full-duplex communication scenarios, a microphone picks up the user's voice along with multiple competing ambient sounds, including background noise, interfering speech and acoustic echoes. It requires efficient acoustic echo cancellation (AEC) and speech enhancement algorithms to extract the target speaker's utterance from the microphone mixture. Any echo leakage or interfering speech leakage could distract the user and degrade the communication experience.


Deep neural network based AEC algorithms flourish with the recent series of AEC-Challenges \cite{sridhar2021icassp,cutler21_interspeech,cutler2022AEC}. 
Westhausen et al.~\cite{westhausen2020acoustic} adopt a dual-signal transformation LSTM network (DTLN) and use both the microphone signal and the far-end signal to predict the near-end speech in an end-to-end fashion. Zhang et al.~\cite{zhang2021ft} propose to use complex neural networks to better model the phase information and frequency-time-LSTM (FT-LSTM) for better temporal modeling. 
Besides the pure neural approaches, hybrid methods that combine neural networks with classical signal processing have shown to generally deliver superior performance. 
Valin et al.~\cite{valin2021low} combine a multi-delay block frequency-domain (MDF) filter and a PercepNet model for joint residual echo suppression and noise reduction. 
Peng et al.~\cite{peng2021acoustic} introduce a time delay estimation (TDE) block besides the adaptive filter to alleviate the task difficulty and propose a gated complex convolutional recurrent neural network (GCCRN) as a post-filter. Zhang et al.~\cite{zhang2022multitask} apply a double-talk friendly weighted recursive least square (wRLS) filter~\cite{wang2021weighted} and multi-task gated convolutional frequency-time-LSTM neural network (GFTNN) to predict the near-end speech activity at the same time. 
Earlier work on DNN-based acoustic echo cancellation also include~\cite{zhang2018deep,seidel21_interspeech}.

\begin{figure}
\centering
    \includegraphics[width=0.45\textwidth]{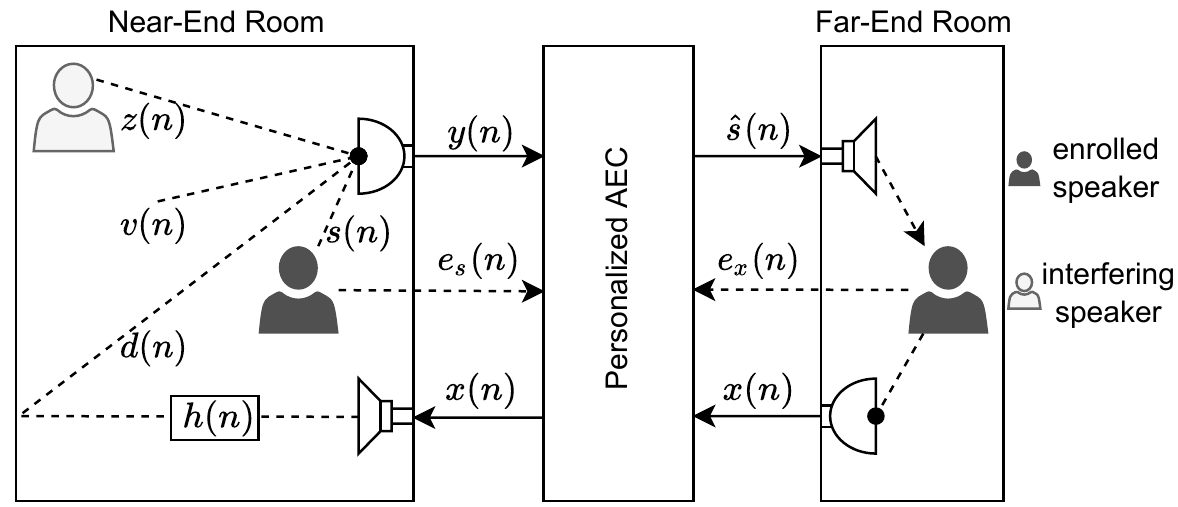}
    \caption{Illustration of personalized AEC in full-duplex communication scenarios.}
    \label{fig:aec_flow}
\end{figure}

\begin{figure*}
\centering
    \includegraphics[width=0.9\textwidth]{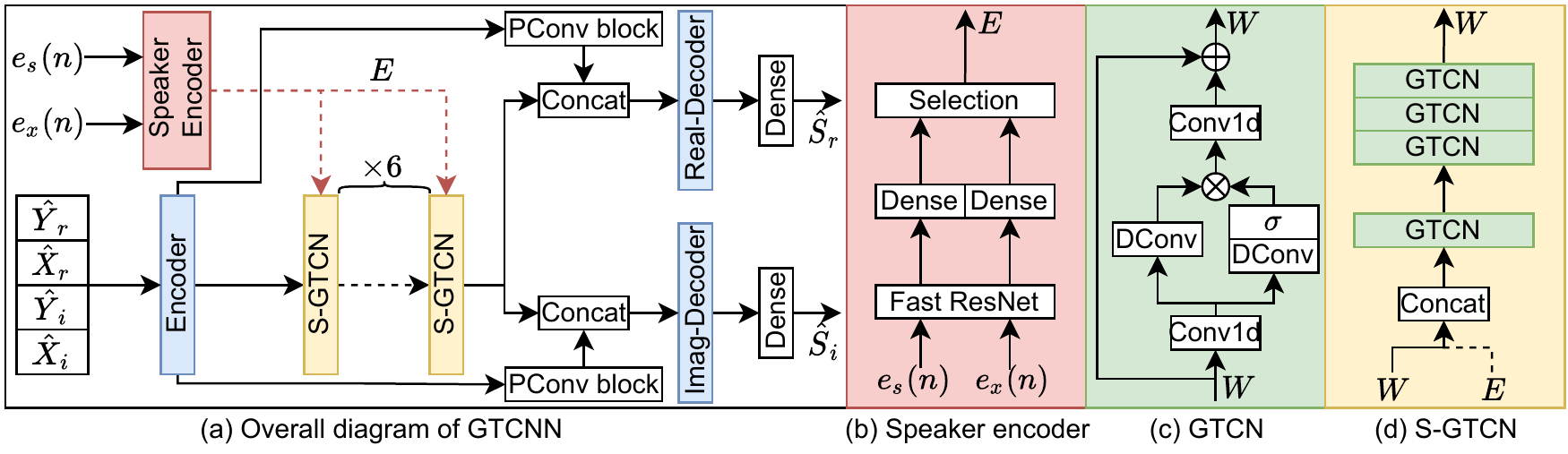}
    \caption{Architecture of the proposed personalized GTCNN. (a) The overall diagram of GTCNN. (b) The speaker encoder. (c) The gated temporal convolution layer (GTCN). (d) The stacked multi-scale GTCN (S-GTCN) block.}
    \label{fig:gtcnn}
\end{figure*}

To the best of our knowledge, current DNN-based AEC models let through all near-end speakers including the interfering speech since there is no auxiliary information for the model to determine whom to attend. This can be undesirable in real-world applications and could potentially lead to privacy issues when the target user shares the space with others~\cite{eskimez2021personalized}. It raises the question of personalized acoustic echo cancellation (PAEC), defined as extracting the target speaker while removing echo, background noise and interfering speech all at once. 
Recently, there has been much attention on personalized speech enhancement (PSE), such as personalized PercepNet~\cite{giri2021personalized} and personalized DCCRN~\cite{eskimez2021personalized}, or target speaker extraction in the field of source separation, such as VoiceFilter~\cite{Wang2019} and SpeakerBeam~\cite{delcroix2020improving}. These methods have explored audio-based~\cite{9067003, 9747765} or visual-based~\cite{gabbay2018visual, wu2019time} clues of the target speaker to filter out interfering speech. Thus PAEC is a natural extension of PSE in full-duplex communication scenarios. 
In PAEC, both parties on the call could have enrolled voice snippets, as shown in Fig.~\ref{fig:aec_flow}. 
A reasonable assumption is that knowing more about the far-end speaker can help echo suppression. Besides, how to make full use of auxiliary information remains an open question and desires an extensive study.

In this paper, we first propose a backbone gated temporal convolutional neural network termed as GTCNN for unconditional AEC. GTCNN is developed on the basis of GCCRN~\cite{peng2021acoustic}, a competitive entry in the INTERSPEECH 2021 AEC-Challenge. By introducing a gated temporal convolution mechanism, GTCNN achieves state-of-the-art performance with fewer parameters.
We then extend GTCNN to the task of PAEC by concatenating speaker embeddings, like d-vectors, of the enrolled users in the hidden layers.
Experiments are conducted in diverse full-duplex scenarios where 
background noise, interfering speech and echo may exist at the same time. It is shown that auxiliary information from either the near-end speaker or the far-end speaker can improve the DNN-based AEC performance.
Nevertheless, ``knowing who is talking" is far more important than ``knowing who you are talking to", and there is still much room for improvement in the utilization of the finite-dimensional speaker embeddings.

\section{Proposed Method}

\subsection{Problem formulation}
\label{sec:formula}
The main architecture of proposed PAEC system is illustrated in Fig.~\ref{fig:aec_flow}.
The microphone signal $y(n)$ consists of the target near-end speech $s(n)$, interfering speech $z(n)$, acoustic echo $d(n)$ and background noise $v(n)$:
\begin{equation}
    y(n) = s(n) + z(n) + d(n) + v(n),
\end{equation}
where $n$ refers to the time sample index. Here $d(n)$ is obtained by convolving the far-end signal $x(n)$ with the acoustic echo path $h(n)$. 
The aim of PAEC is to recover the target speaker's voice from the microphone mixture, and this is achieved by:
\begin{equation}
    \hat{s}(n) = \text{DNN}(y(n), x(n); e_s(n), e_x(n)),
\end{equation}
where $e_s(n)$ and $e_x(n)$ are speech snippets enrolled from the near-end speaker and the far-end speaker, respectively.

\subsection{Feature extraction}

The proposed method works on the complex spectrum domain.
Window length and hop size are 20 ms and 10 ms, respectively. Then a 320-point short-time Fourier transform (STFT) is applied to each frame of $y(n)$ and $x(n)$ to produce the complex spectra $Y(t, f)$ and $X(t, f)$, respectively. The input features are the real and imaginary parts of the compressed spectrum:
\begin{equation}
\label{compress}
\begin{aligned}
\hat{Y}_{r}(t, f) &=|Y(t, f)|^{0.5} \cos (\angle Y(t, f)) \\
\hat{Y}_{i}(t, f) &=|Y(t, f)|^{0.5} \sin (\angle Y(t, f))
\end{aligned},
\end{equation}
where $\angle Y(t, f)$ is the phase of the microphone signal, and $\hat{X}_{r}(t, f), \hat{X}_{i}(t, f)$ are obtained in the same way. The compressed feature has proved to be helpful for the AEC task~\cite{peng2021acoustic}. Finally, $\hat{Y}_{r/i}(t, f)$ and $\hat{X}_{r/i}(t, f)$ are stacked together on the channel axis. 

\subsection{Speaker encoder}

Speaker embedding is used to identify the target speaker in the observed signal to distinguish the target speech from background noise, interfering speech and echo. In the proposed PAEC system, a lightweight speaker recognition neural network termed as fast ResNet~\cite{chung2020in}, is used to extract speaker embeddings from speech snippets enrolled from the near-end and far-end users. Note that other speaker embedding networks can be easily adopted as well. The fast ResNet can extract speaker embeddings with small intra-speaker and large inter-speaker distances in low computational complexity. We use pre-trained fast ResNet provided by~\cite{chung2020in}, denoted as $\mathcal{E}$, and  cascade a dense layer $\mathcal{D}$ after to project the 512-dimensional utterance-level speaker embeddings into more compact 256-dimensional representations:

\begin{equation}
\label{se}
\begin{aligned}
E_s &= \mathcal{D}_s \left( \mathcal{E}\left(e_{s}(n) ; \boldsymbol{\Phi}_{\mathcal{E}}\right) ; \boldsymbol{\Phi}_{\mathcal{D}_s} \right) \\
E_x &= \mathcal{D}_x \left( \mathcal{E}\left(e_{x}(n) ; \boldsymbol{\Phi}_{\mathcal{E}}\right) ; \boldsymbol{\Phi}_{\mathcal{D}_x} \right)
\end{aligned},
\end{equation}
where $\mathcal{D}_s$ and $\mathcal{D}_x$ denote the projection layers for $e_{s}(n)$ and $e_{x}(n)$, respectively. $\boldsymbol{\Phi}_{\mathcal{E}}$, $\boldsymbol{\Phi}_{\mathcal{D}_s}$, $\boldsymbol{\Phi}_{\mathcal{D}_x}$ are the corresponding parameter sets. The utterance-level speaker representation $E_s$ and $E_x$ will be repeated along the time axis during training, to make their lengths consistent with the input signals.
The resulting repeated feature is of shape $\mathbb{R}^{T\times 256}$, with $T$ the number of time-frames. 
The \textit{Selection} operation in Fig.~\ref{fig:gtcnn}(b) is used to determine the combination of different speaker embeddings. The final output of speaker encoder is represented by $E \in \{None, E_s, E_x, E_{\text{mix}} \}$, where $None$ means no  auxiliary information is used, and $E_{\text{mix}} = [E_s; E_x] \in \mathbb{R}^{T \times 512}$.

\subsection{GTCNN}

As illustrated in Fig.~\ref{fig:gtcnn}, the proposed GTCNN consists of three modules -- \textbf{encoder} for feature extraction, \textbf{decoder} for estimated spectra reconstruction and \textbf{stacked GTCN blocks} for sequence modeling. The speaker embeddings are concatenated in the S-GTCN blocks as in Fig.~\ref{fig:gtcnn}(d).

The {encoder} consists of 5 gated Conv-2d (GConv) layers, and the {real-decoder} and {imag-decoder} both consist of 5 gated transposed Conv-2d (TrGConv) layers, which are used to reconstruct the real and imaginary parts of the estimated speech spectrum, respectively. More details of GConv and TrGConv can be found in~\cite{zhang2022multitask}. Between the corresponding GConv layer and TrGConv layer, a pointwise Conv-2d (PConv) layer is used to connect the shallow and deep feature representations.

In the GTCN layer as illustrated in Fig.~\ref{fig:gtcnn}(c), the first PConv layer is used to compress the feature dimension and the last PConv layer is used to restore the feature dimension as with the input. Dilated Conv-1d (DConv) layers are inserted in between to expand the receptive field~\cite{bai2018empirical}. 
The convolution kernel size of DConv layers is set to 3. 
$\sigma$ denotes the sigmoid function, $\otimes$ denotes point-wise multiplication and $\oplus$ denotes the residual connection between the input and output.
The gate mechanism~\cite{tan2018gated} is introduced to control the information flow dynamically. 
PReLU~\cite{he2015delving} activation and instance normalization~\cite{ulyanov2016instance} are applied before each DConv layer.

The S-GTCN block, as illustrated in Fig.~\ref{fig:gtcnn}(d), consists of four stacked GTCN layers with different receptive fields. The dilations are set to 1, 2, 5, and 9, respectively. Such stacked structure utilizes up to 34 frames of historical information. In the experiments, using S-GTCN for sequence modeling outperforms a short-term memory (LSTM) alternative.
Note that the dotted lines in Fig.~\ref{fig:gtcnn}(a) and (d) are optional in different configurations. The speaker embeddings, if used, will be concatenated with the input of all S-GTCN blocks at the feature dimension. 

When $E = None$, GTCNN works as an unconditional AEC model. For $E = E_s$ or $E = E_x$, the near-end enrolled speech or far-end enrolled speech is used as auxiliary information, termed as personalized GTCNN-$E_s$ (pGTCNN-$E_s$) and pGTCNN-$E_x$, respectively. For $E = E_{\text{mix}}$, enrolled speech from both ends is used as auxiliary information, termed as pGTCNN-$E_\text{mix}$. The three personalized models only differ in the input feature size.
Detailed input/output configurations of each module are explained in Table~\ref{config}.

\begin{table}[]
\caption{Configuration of the proposed GTCNN, taking pGTCNN-$E_s$ or pGTCNN-$E_x$ as an example.}
\label{config}
\centering

\begin{tabular}[htb]{lll}
\toprule 
\multicolumn{1}{l}{{Module Name}} & \multicolumn{1}{l}{{Input Size}} & \multicolumn{1}{l}{{Output Size}} \\
\midrule
Encoder & $\text{T} \times \text{4} \times \text{161}$ & $\text{T} \times \text{C} \times \text{4}$ \\ 
Speaker encoder & $e(n)$ & $\text{T} \times \text{256} $ \\ 
S-GTCN concat & 
\makecell[l]{$\text{T} \times \text{C} \times \text{4}$ \\ $\text{T} \times \text{256}$}
 & $\text{T} \times \text{(4C + 256)}$ \\ 
S-GTCN & $\text{T} \times \text{(4C + 256)}$ & $\text{T} \times \text{4C}$ \\ 
S-GTCN reshape & $\text{T} \times \text{4C}$ & $\text{T} \times \text{C} \times \text{4}$ \\ 
Real/Imag-decoder & $\text{T} \times \text{C} \times \text{4}$  & $\text{T} \times \text{1} \times \text{161}$ \\ 
Dense & $\text{T} \times \text{1} \times \text{161}$ & $\text{T} \times \text{1} \times \text{161}$ \\ 
\bottomrule
\end{tabular}
\end{table}

\subsection{Cost function}
\label{ssec:cost}

The output $W(t, f)$ of the last dense layers of the proposed GTCNN model are decompressed as follow:
\begin{equation}
\label{uncompress}
\begin{aligned}
\hat{S}_{r}(t, f) &=|W(t, f)|^{2} \cos (\angle W(t, f)) \\
\hat{S}_{i}(t, f) &=|W(t, f)|^{2} \sin (\angle W(t, f))
\end{aligned},
\end{equation}
where $\hat{S}_r(t, f)$ and $\hat{S}_i(t, f)$ are interpreted as the real and imaginary parts of the estimated complex spectra of $\hat{s}(n)$, respectively.
The cost function $\mathcal{L}$ used in all experiments is given by:
\begin{equation}
\begin{aligned}
\mathcal{L}_{\text {r}}(t, f) &= |S_r(t, f)- \hat{S}_r(t, f)|^{2} \\
\mathcal{L}_{\text {i}}(t, f) &= | S_i(t, f)- \hat{S}_i(t, f)|^{2} \\
\mathcal{L} &= \frac{1}{T}\frac{1}{F}\sum_{T}\sum_{F}\left(\mathcal{L}_{\text {r}}(t, f) + \mathcal{L}_{\text {i}}(t, f)\right)
\end{aligned},
\end{equation}
which allows both real and imaginary parts to be optimized at the same time.

\begin{table*}[!htbp]
\footnotesize
\centering

\setlength{\tabcolsep}{4pt}
\caption{Comparison of the personalized AEC models under different scenarios. DT: double talk, ST: single-talk, NE: near-end, FE: far-end. Bold results indicate the best in each column and the overall best results are underlined. }
\label{metric1}
\begin{tabular}{@{\extracolsep{4pt}}lcccccccccccccc@{}}
\toprule
 Method  & \#Params(M) & \multicolumn{8}{c}{DT-Noise}  & \multicolumn{2}{c}{ST-NE} &  \multicolumn{2}{c}{ST-FE} & \multirow{4}{*}{Data}\\
\cline{1-2}  \cline{ 3-10}  \cline{ 11-12 }  \cline{13-14 }
\multicolumn{2}{c}{SER (in dB)}&
\multicolumn{2}{c}{5} &
\multicolumn{2}{c}{15} &
\multicolumn{2}{c}{5} &
\multicolumn{2}{c}{15} &
\multicolumn{2}{c}{+$\infty$} &
\multicolumn{2}{c}{-} & \\
\cline{1-2}  \cline{ 3-6} \cline{ 7-10}  \cline{ 11-12 }  \cline{13-14 }
\multicolumn{2}{c}{SIR (in dB)} &
5 &
15 &
5  &
15  &
5  &
15  &
5  &
15  &
15  &
+$\infty$  &
15  &
+$\infty$  & \\
\cline{1-2}  \cline{ 3-6} \cline{ 7-10}  \cline{ 11-12 }  \cline{13-14 }
\multicolumn{2}{c}{Metric} & \multicolumn{4}{c}{WER$\downarrow$} &\multicolumn{4}{c}{PESQ$\uparrow$} &\multicolumn{2}{c}{PESQ$\uparrow$} &\multicolumn{2}{c}{ERLE$\uparrow$} \\
\hline
Input&- &61.16 & 38.27&\textbf{32.72} &\underline{\textbf{12.70}} &1.91&2.12&2.11&2.47&2.87&4.50&\multicolumn{2}{c}{0}&- \cr 
DTLN~\cite{westhausen2020acoustic}&10.40 & 52.64&20.76 &46.63 & 16.64 &2.05&2.52&2.12&2.68&2.88&4.19&15.65&32.06 & - \cr
GCCRN~\cite{peng2021acoustic}&10.20 &54.18 & 20.60&44.84 &15.87 &2.01&2.51&2.12&2.75&2.96&4.22&17.54&39.98 & \multirow{6}{*}{D1}\cr
GTCNN&3.26 & 52.55& 19.10&44.41 &13.89 &2.05&2.57&2.17&2.78&3.01&4.22&19.23&43.41 \cr 
pGTCNN-$E_s$&3.36 &\textbf{47.20} &\underline{\textbf{17.67}} &38.13 &13.48 &\textbf{2.08}&\underline{\textbf{2.64}}&\textbf{2.20}&\underline{\textbf{2.84}}&\textbf{3.03}&\underline{\textbf{4.28}}&19.68&\underline{\textbf{46.91}} \cr 
pGTCNN-$E_x$&3.36 &50.80 & 18.67& 39.41&14.44 &2.07&2.62&2.18&2.82&3.03&4.27&18.24&46.38 \cr 
pGTCNN-$E_\text{mix}$&3.46 &49.61 & 18.53&41.00 &13.53 &\textbf{2.08}&2.62&2.19&2.82&2.99&4.24&19.60&43.81 \cr
GTCNN-L&3.47 & 50.52&18.94 &42.97 &14.71 &2.06&2.59&2.17&2.80&2.99&4.21&\textbf{21.19}&43.65 \cr 
\hline
pGTCNN-$E_s$&3.36 &\underline{\textbf{34.78}} & \textbf{19.65}& \textbf{32.31}&\textbf{15.01} &\textbf{2.18}&\textbf{2.58}&\textbf{2.36}&\textbf{2.83}&\textbf{3.15}&\textbf{4.11}&\textbf{40.80}&\textbf{42.82} &\multirow{4}{*}{D2}\cr 
pGTCNN-$E_x$&3.36 & 37.16& 20.55& 35.30&15.76 &2.14&2.53&2.30&2.79&3.08&4.07&37.74&42.21 \cr 
pGTCNN-$E_\text{mix}$&3.46 &36.07 & 19.80& 34.14&15.83 &\textbf{2.18}&2.55&\textbf{2.36}&2.81&3.12&4.10&39.10&40.91 \cr 
GTCNN-L&3.47 & 41.56& 21.87&37.40 &16.37 &2.08&2.53&2.27&2.72&3.03&4.02&30.54&40.91\cr
\hline
pGTCNN-$E_s$&3.36 & \textbf{39.97}& \textbf{18.37}& \underline{\textbf{28.54}}&\textbf{13.39} &\underline{\textbf{2.20}}&\textbf{2.60}&\underline{\textbf{2.37}}&\underline{\textbf{2.84}}&\underline{\textbf{3.20}}&\textbf{4.15}&\underline{\textbf{41.61}}&\textbf{43.53} &\multirow{4}{*}{D3}\cr 
pGTCNN-$E_x$&3.36 &43.72 &20.46 &29.99 &14.39 &2.15&2.53&2.32&2.79&3.15&4.13&38.22&41.62 \cr 
pGTCNN-$E_\text{mix}$&3.46 &41.20 & 18.87&30.83 & 13.35&2.19&2.56&2.36&2.82&3.18&4.14&39.77&41.12 \cr
GTCNN-L&3.47 &46.13 &21.01 &33.77 &14.94 &2.10&2.51&2.27&2.75&3.11&4.04&37.89&40.48\cr
\bottomrule
\end{tabular}
\end{table*}

\section{Experiments}
\label{sec:exp}

\subsection{Datasets}
\label{ssec:dataset}

We use the LibriSpeech~\cite{panayotov2015librispeech} corpus to generate the training and validation datasets. 
The near-end speech $s(n)$, far-end speech $x(n)$ and optional interfering speech $z(n)$ are randomly chosen from the corpus while ensuring that they are not from the same speaker. 
We simulate rooms of different sizes with $width \times height \times depth$, where $width$ is uniformly distributed in $[5, 8]$ m, $height$ in $[3, 4]$ m and $depth$ in $[3, 5]$ m.
Room impulse responses (RIRs) $h(n)$ are simulated by the HYB method described in~\cite{bezzam2020study}. The RT60 is set to fall within the range of $[0.2, 0.7]$ s. The far-end speech $x(n)$ is convoluted with $h(n)$ to generate the echo signal $d(n)$. A random delay between 0 and 512 ms is applied as the time delay between playback and recording to $d(n)$.
Background noise files in the training set are drawn from the deep noise suppression (DNS) challenge noise set~\cite{reddy2021interspeech}.
The microphone signal is then obtained by mixing $s(n)$, $z(n)$, $d(n)$ and $v(n)$ according to the preset signal-to-interference ratio (SIR), signal-to-echo ratio (SER) and signal-to-noise ratio (SNR), where SIR is defined as:
\begin{equation}
\label{sir}
    \text{SIR}=10 \log _{10}\left(\sum_{n} s^{2}(n) / \sum_{n} z^{2}(n)\right),
\end{equation}
and SER, SNR are defined by replacing $z(n)$ in Eq.~(\ref{sir}) with $d(n)$ and $v(n)$, respectively. We empirically assume that the energy of interfering speaker is always less than that of target speaker (i.e. $\text{SIR} \geq 0$).

We design three scenarios to compare the model's performance with different training datasets: 
\begin{itemize}
    \item $\text{SIR}=+\infty$ (i.e. $z(n)=0$), $\text{SER} \in [-10, 20]$ and $\text{SNR} \in [-5, 40]$, for a general scenario where not any interfering speaker is considered;
    \item $\text{SIR} \in [0, 20]$, $\text{SER} \in [-10, 20]$ and $\text{SNR} \in [-5, 40]$, for an adverse scenario where background noise, interfering speaker and echo exist at the same time;
    \item $\text{SIR} \in [0, 20]$, $\text{SER} \in [-10, 20]$ and $\text{SNR} \in [15, 45]$, for a mild scenario where the proportion of background noise is reduced, and the focus is echo cancellation and interfering speech suppression.
\end{itemize}
For brevity, we name the three scenarios in the following as “D1”, “D2”, and “D3”, respectively.

There are 200 hours of training data, and 20 hours of validation data for each scenario. The datasets cover all 2,338 speakers in the corpus. For each speaker, a fixed 60-second utterance is selected and adopted as the enrolled speech.

We use the voice cloning toolkit (VCTK) corpus~\cite{yamagishi2019vctk} as clean speech and the DEMAND~\cite{thiemann2013diverse} dataset as noise for creating the testset. 
The signals are first downsampled to 16k Hz.
It should be noted that files with similar audio lengths in VCTK are likely to have the same speech content. 
So we make sure that the audio lengths of the near-end speakers, far-end speakers and interference speakers are also different. 
Finally, there are 140 utterances longer than 10 s for each scenario (with different SIRs, SERs and SNRs) generated as the test set.

\subsection{Performance metrics}
\label{ssec:metric2}

The echo suppression performance is evaluated in terms of echo return loss enhancement (ERLE), which is defined as 
\begin{equation}
\label{erle}
\mathrm{ERLE}=10\log_{10}\left[\sum_{n} y^{2}(n) / \sum_{n} \hat{s}^{2}(n)\right].
\end{equation}
Meanwhile, we use perceptual evaluation of speech quality (PESQ)~\cite{rix2001perceptual} for double-talk (DT) and near-end single-talk (ST) periods. The higher the scores, the better the performance. For reference, word error rate (WER) is also calculated for the DT periods, which is well related to the interference suppression performance.

\subsection{Implementation details}
\label{ssec:expsettings}

In GTCNN, the convolution kernel size of each GConv and TrGConv layer is (2, 3), and the stride size is (1, 2). To maintain a causal model, the inputs of all convolutional layers are left padded with zeros.
The output channel $C$ is set to 80, except for the last TrGConv layer, $C=1$ , which is used to estimate the real/imaginary part of the clean near-end spectra. The output channel of Conv1d is set to 64.
For the ablation study, a GTCNN-L model of the same large parameter size is included for a fair comparison. 

All neural models are trained with the Adam optimizer~\cite{kingma2014adam} with an initial learning rate of 1e-4. The learning rate is halved if there is no loss decrease on validation set for 2 epochs. 

\subsection{Results and analysis}
\label{ssec:performace}

Table~\ref{metric1} presents the comparison of GTCNN using auxiliary information under different training datasets. DT-Noise means adding noise with $\text{SNR} = 10$ dB. The best results in each column and the best overall results are bolded and underlined, respectively. 
Samples of the processed audio clips can be found on our demo page\footnote{https://echocatzh.github.io/GTCNN}.

From Table~\ref{metric1}, we first see that GTCNN, with fewer parameters and the introduced gated mechanism, outperforms GCCRN and an open-source DTLN~\cite{westhausen2020acoustic} in all situations. Comparing unconditional GTCNN with its personalized variants pGTCNN-$E_s$, pGTCNN-$E_x$ and pGTCNN-$E_\text{mix}$, we find that the auxiliary information enrolled from either the near-end speaker or the far-end speaker can improve PESQ and ERLE scores. 
Overall, GTCNN-$E_s$ outperforms pGTCNN-$E_x$ in all three scenarios (D1, D2, D3). Obviously, ``knowing who is talking" is far more important than ``knowing who you are talking to". 
Another reason is that the far-end speech is affected by the varying acoustic echo paths so that $E_x$ can not fully characterize the far-end speaker in the mixture signal. 
As shown in ST-FE scenario with SIR $= 15$ dB, explicitly adding interfering speech during training can improve interference suppression in the test set even for models without auxiliary information (comparing D1 with D2/D3). 
However, based on the results in the ST-NE scenario with SIR$= +\infty$, near-end speech distortion is introduced at the same time. 
The overall performance of models trained with D3 is better than that of the models trained with D2. It indicates that a too complicated training data setup would sometimes mess up the model training process. Similar findings are also reported in ~\cite{fu2021desnet}.

The results also indicate that the considered auxiliary information is still insufficient for interference suppression. Comparing the (SER=5 dB, SIR=15 dB) column with the (SER=15 dB, SIR=5 dB) column, the reduction in WER is obviously higher, though the target-to-non-target energy ratios are the same. The acoustic echoes can be effectively suppressed since the far-end reference signal is used as input, while not any information of the interference is given to the model.
The one that does not match expectations is that pGTCNN-$E_\text{mix}$ performs no better than pGTCNN-$E_x$, though more information is provided. This could attribute to the confusion of the speaker information provided by both the near-end and the far-end speakers. 
As has been pointed out in~\cite{elminshawi2022new}, the finite-dimensional speaker embeddings may not have been efficiently utilized in personalized models.

\section{Conclusions}
\label{sec:end}
In this paper, we have studied the personalized acoustic echo cancellation problem for full-duplex communications, where interfering speech, echo and background noise exist at the same time. In conclusion, the proposed GTCNN outperforms state-of-the-art AEC models, and auxiliary information is indeed helpful for the model, but there is still room for improvement. In future work, we will test whether more advanced speaker recognition models with lower equal error rates will provide consistent performance gain for the AEC models, and explore more effective ways to utilize the speaker embeddings.

\bibliographystyle{IEEEtran}

\bibliography{refs}

\begin{thebibliography}{10}
\providecommand{\url}[1]{#1}
\csname url@samestyle\endcsname
\providecommand{\newblock}{\relax}
\providecommand{\bibinfo}[2]{#2}
\providecommand{\BIBentrySTDinterwordspacing}{\spaceskip=0pt\relax}
\providecommand{\BIBentryALTinterwordstretchfactor}{4}
\providecommand{\BIBentryALTinterwordspacing}{\spaceskip=\fontdimen2\font plus
\BIBentryALTinterwordstretchfactor\fontdimen3\font minus
  \fontdimen4\font\relax}
\providecommand{\BIBforeignlanguage}[2]{{%
\expandafter\ifx\csname l@#1\endcsname\relax
\typeout{** WARNING: IEEEtran.bst: No hyphenation pattern has been}%
\typeout{** loaded for the language `#1'. Using the pattern for}%
\typeout{** the default language instead.}%
\else
\language=\csname l@#1\endcsname
\fi
#2}}
\providecommand{\BIBdecl}{\relax}
\BIBdecl

\bibitem{sridhar2021icassp}
K.~Sridhar, R.~Cutler, A.~Saabas, T.~Parnamaa, M.~Loide, H.~Gamper, S.~Braun,
  R.~Aichner, and S.~Srinivasan, ``{ICASSP} 2021 acoustic echo cancellation
  challenge: Datasets, testing framework, and results,'' in
  \emph{ICASSP}.\hskip 1em plus 0.5em minus 0.4em\relax IEEE, 2021, pp.
  151--155.

\bibitem{cutler21_interspeech}
R.~Cutler, A.~Saabas, T.~Parnamaa, M.~Loide, S.~Sootla, M.~Purin, H.~Gamper,
  S.~Braun, K.~Sorensen, R.~Aichner, and S.~Srinivasan, ``{INTERSPEECH 2021}
  acoustic echo cancellation challenge,'' in \emph{INTERSPEECH}, 2021, pp.
  4748--4752.

\bibitem{cutler2022AEC}
R.~Cutler, A.~Saabas, T.~Parnamaa, M.~Purin, H.~Gamper, S.~Braun, K.~Sorensen,
  and R.~Aichner, ``{ICASSP} 2022 acoustic echo cancellation challenge,'' in
  \emph{ICASSP}.\hskip 1em plus 0.5em minus 0.4em\relax IEEE, 2022.

\bibitem{westhausen2020acoustic}
N.~Westhausen and B.~Meyer, ``Acoustic echo cancellation with the dual-signal
  transformation {LSTM} network,'' in \emph{ICASSP}.\hskip 1em plus 0.5em minus
  0.4em\relax IEEE, 2021, pp. 7138--7142.

\bibitem{zhang2021ft}
S.~Zhang, Y.~Kong, S.~Lv, Y.~Hu, and L.~Xie, ``{F-T-LSTM} based complex network
  for joint acoustic echo cancellation and speech enhancement,'' in
  \emph{INTERSPEECH}.\hskip 1em plus 0.5em minus 0.4em\relax ISCA, 2021, pp.
  4758--4762.

\bibitem{valin2021low}
J.-M. Valin, S.~Tenneti, K.~Helwani, U.~Isik, and A.~Krishnaswamy,
  ``Low-complexity, real-time joint neural echo control and speech enhancement
  based on percepnet,'' in \emph{ICASSP}.\hskip 1em plus 0.5em minus
  0.4em\relax IEEE, 2021, pp. 7133--7137.

\bibitem{peng2021acoustic}
R.~Peng, L.~Cheng, C.~Zheng, and X.~Li, ``Acoustic echo cancellation using deep
  complex neural network with nonlinear magnitude compression and phase
  information,'' \emph{INTERSPEECH}, pp. 4768--4772, 2021.

\bibitem{zhang2022multitask}
S.~Zhang, Z.~Wang, J.~Sun, Y.~Fu, B.~Tian, Q.~Fu, and L.~Xie, ``Multi-task deep
  residual echo suppression with echo-aware loss,'' in \emph{ICASSP}.\hskip 1em
  plus 0.5em minus 0.4em\relax IEEE, 2022.

\bibitem{wang2021weighted}
Z.~Wang, Y.~Na, Z.~Liu, B.~Tian, and Q.~Fu, ``Weighted recursive least square
  filter and neural network based residual echo suppression for the
  aec-challenge,'' in \emph{ICASSP}.\hskip 1em plus 0.5em minus 0.4em\relax
  IEEE, 2021, pp. 141--145.

\bibitem{zhang2018deep}
H.~Zhang and D.~Wang, ``Deep learning for acoustic echo cancellation in noisy
  and double-talk scenarios,'' \emph{INTERSPEECH}, pp. 3239--3243, 2018.

\bibitem{seidel21_interspeech}
E.~Seidel, J.~Franzen, M.~Strake, and T.~Fingscheidt, ``Y2-net {FCRN} for
  acoustic echo and noise suppression,'' in \emph{INTERSPEECH}, 2021, pp.
  4763--4767.

\bibitem{eskimez2021personalized}
S.~E. Eskimez, T.~Yoshioka, H.~Wang, X.~Wang, Z.~Chen, and X.~Huang,
  ``Personalized speech enhancement: New models and comprehensive evaluation,''
  2022.

\bibitem{giri2021personalized}
R.~Giri, S.~Venkataramani, J.-M. Valin, U.~Isik, and A.~Krishnaswamy,
  ``Personalized percepnet: Real-time, low-complexity target voice separation
  and enhancement,'' pp. 1124--1128, 2021.

\bibitem{Wang2019}
Q.~Wang, H.~Muckenhirn, K.~Wilson, P.~Sridhar, Z.~Wu, J.~R. Hershey, R.~A.
  Saurous, R.~J. Weiss, Y.~Jia, and I.~L. Moreno, ``{VoiceFilter: Targeted
  Voice Separation by Speaker-Conditioned Spectrogram Masking},'' in
  \emph{INTERSPEECH}, 2019, pp. 2728--2732.

\bibitem{delcroix2020improving}
M.~Delcroix, T.~Ochiai, K.~Zmolikova, K.~Kinoshita, N.~Tawara, T.~Nakatani, and
  S.~Araki, ``Improving speaker discrimination of target speech extraction with
  time-domain speakerbeam,'' in \emph{ICASSP}.\hskip 1em plus 0.5em minus
  0.4em\relax IEEE, 2020, pp. 691--695.

\bibitem{9067003}
C.~Xu, W.~Rao, E.~S. Chng, and H.~Li, ``Spex: Multi-scale time domain speaker
  extraction network,'' \emph{IEEE Trans. Signal Process.}, vol.~28, pp.
  1370--1384, 2020.

\bibitem{9747765}
Y.~Ju, W.~Rao, X.~Yan, Y.~Fu, S.~Lv, L.~Cheng, Y.~Wang, L.~Xie, and S.~Shang,
  ``{TEA-PSE: Tencent-Ethereal-Audio-Lab Personalized Speech Enhancement System
  for ICASSP 2022 DNS Challenge},'' in \emph{ICASSP}, 2022, pp. 9291--9295.

\bibitem{gabbay2018visual}
A.~Gabbay, A.~Shamir, and S.~Peleg, ``Visual speech enhancement,''
  \emph{INTERSPEECH}, pp. 1170--1174, 2018.

\bibitem{wu2019time}
J.~Wu, Y.~Xu, S.-X. Zhang, L.-W. Chen, M.~Yu, L.~Xie, and D.~Yu, ``Time domain
  audio visual speech separation,'' in \emph{ASRU}.\hskip 1em plus 0.5em minus
  0.4em\relax IEEE, 2019, pp. 667--673.

\bibitem{chung2020in}
J.~S. Chung, J.~Huh, S.~Mun, M.~Lee, H.~S. Heo, S.~Choe, C.~Ham, S.~Jung, B.-J.
  Lee, and I.~Han, ``In defence of metric learning for speaker recognition,''
  in \emph{INTERSPEECH}, 2020.

\bibitem{bai2018empirical}
S.~Bai, J.~Z. Kolter, and V.~Koltun, ``An empirical evaluation of generic
  convolutional and recurrent networks for sequence modeling,'' \emph{arXiv
  preprint arXiv:1803.01271}, 2018.

\bibitem{tan2018gated}
K.~Tan, J.~Chen, and D.~Wang, ``Gated residual networks with dilated
  convolutions for monaural speech enhancement,'' \emph{IEEE Trans. Signal
  Process.}, vol.~27, no.~1, pp. 189--198, 2018.

\bibitem{he2015delving}
K.~He, X.~Zhang, S.~Ren, and J.~Sun, ``Delving deep into rectifiers: Surpassing
  human-level performance on imagenet classification,'' in \emph{ICCV}, 2015,
  pp. 1026--1034.

\bibitem{ulyanov2016instance}
D.~Ulyanov, A.~Vedaldi, and V.~Lempitsky, ``Instance normalization: The missing
  ingredient for fast stylization,'' \emph{arXiv preprint arXiv:1607.08022},
  2016.

\bibitem{panayotov2015librispeech}
V.~Panayotov, G.~Chen, D.~Povey, and S.~Khudanpur, ``Librispeech: an {ASR}
  corpus based on public domain audio books,'' in \emph{ICASSP}.\hskip 1em plus
  0.5em minus 0.4em\relax IEEE, 2015, pp. 5206--5210.

\bibitem{bezzam2020study}
E.~Bezzam, R.~Scheibler, C.~Cadoux, and T.~Gisselbrecht, ``A study on more
  realistic room simulation for far-field keyword spotting,'' in \emph{APSIPA
  ASC}.\hskip 1em plus 0.5em minus 0.4em\relax IEEE, 2020, pp. 674--680.

\bibitem{reddy2021interspeech}
C.~K. Reddy, H.~Dubey, K.~Koishida, A.~Nair, V.~Gopal, R.~Cutler, S.~Braun,
  H.~Gamper, R.~Aichner, and S.~Srinivasan, ``Interspeech 2021 deep noise
  suppression challenge,'' \emph{INTERSPEECH}, pp. 2796--2800, 2021.

\bibitem{yamagishi2019vctk}
J.~Yamagishi, C.~Veaux, and K.~MacDonald, ``{CSTR VCTK Corpus}: English
  multi-speaker corpus for {CSTR} voice cloning toolkit (version 0.92),'' 2019.

\bibitem{thiemann2013diverse}
J.~Thiemann, N.~Ito, and E.~Vincent, ``The diverse environments multi-channel
  acoustic noise database (demand): A database of multichannel environmental
  noise recordings,'' in \emph{POMA}, vol.~19, no.~1.\hskip 1em plus 0.5em
  minus 0.4em\relax ASA, 2013, p. 035081.

\bibitem{rix2001perceptual}
A.~W. Rix, J.~Beerends, M.~Hollier, and A.~Hekstra, ``Perceptual evaluation of
  speech quality (pesq)-a new method for speech quality assessment of telephone
  networks and codecs,'' in \emph{ICASSP}, vol.~2.\hskip 1em plus 0.5em minus
  0.4em\relax IEEE, 2001, pp. 749--752.

\bibitem{kingma2014adam}
\BIBentryALTinterwordspacing
D.~Kingma and J.~B., ``Adam: {A} method for stochastic optimization,'' in
  \emph{ICLR}, Y.~Bengio and Y.~LeCun, Eds., 2015. [Online]. Available:
  \url{http://arxiv.org/abs/1412.6980}
\BIBentrySTDinterwordspacing

\bibitem{fu2021desnet}
Y.~Fu, J.~Wu, Y.~Hu, M.~Xing, and L.~Xie, ``{DESNet}: A multi-channel network
  for simultaneous speech dereverberation, enhancement and separation,'' in
  \emph{SLT}.\hskip 1em plus 0.5em minus 0.4em\relax IEEE, 2021, pp. 857--864.

\bibitem{elminshawi2022new}
M.~Elminshawi, W.~Mack, S.~Chakrabarty, and E.~A. Habets, ``New insights on
  target speaker extraction,'' \emph{arXiv preprint arXiv:2202.00733}, 2022.

\end{thebibliography}

\end{document}